\documentclass[prl,showpacs,showkeys,twocolumn,nofootinbib]{revtex4}
\usepackage{graphicx}
\usepackage{bm}
\begin{document}
\title{TeV Astrophysics Constraints on Planck Scale Lorentz Violation}
\author{Ted Jacobson}
\email{jacobson@physics.umd.edu}
\affiliation{Department of Physics, University of Maryland,
College Park, MD 20742--4111, USA}
\author{Stefano Liberati}
\email{liberati@physics.umd.edu}
\affiliation{Department of Physics, University of Maryland,
College Park, MD 20742--4111, USA}
\author{David Mattingly}
\email{davemm@physics.umd.edu}
\affiliation{Department of Physics, University of Maryland,
College Park, MD 20742--4111, USA}
\date{14 December 2001; \LaTeX-ed \today}
\bigskip
\begin{abstract}
We analyze observational constraints from TeV astrophysics on Lorentz
violating nonlinear dispersion for photons and electrons without
assuming any {\em a priori} equality between the photon and electron
parameters. The constraints arise from thresholds for vacuum
\v{C}erenkov radiation, photon decay and photo-production of
electron-positron pairs.  We show that the parameter plane for cubic
momentum terms in the dispersion relations is constrained to an order
unity region in Planck units. We find that the threshold configuration
can occur with an asymmetric distribution of momentum for pair
creation, and with a hard photon for vacuum \v{C}erenkov radiation.
\end{abstract}
\maketitle
\def\half{{1\over2}}
\def\L{{\mathcal L}}
\def\S{{\mathcal S}}
\def\d{{\mathrm{d}}}
\def\etal{{\emph{et al.}}}
\def\det{{\mathrm{det}}}
\def\tr{{\mathrm{tr}}}
\def\ie{{\emph{i.e.}}}
\def\eg{{\emph{e.g.}}}
\def\im{{\rm i}}
\def\bnabla{\mbox{\boldmath$\nabla$}}
\def\x{{\mathbf x}}
\def\aka{{\emph{aka}}}
\def\Choose#1#2{{#1 \choose #2}}
\def\etc{{\emph{etc.}}}
\def\gsim{\; \raisebox{-.8ex}{$\stackrel{\textstyle >}{\sim}$}\;}
\def\lsim{\; \raisebox{-.8ex}{$\stackrel{\textstyle <}{\sim}$}\;}
\def\st{\scriptstyle}
\def\sst{\scriptscriptstyle}
\def\mco{\multicolumn}
\def\epp{\epsilon^{\prime}}
\def\vep{\varepsilon}
\def\ra{\rightarrow}
\def\ppg{\pi^+\pi^-\gamma}
\def\vp{{\bf p}}
\def\al{\alpha}
\def\ab{\bar{\alpha}}
\def\half{{1\over2}}
\def\L{{\mathcal L}}
\def\d{{\mathrm{d}}}
\def\p{{\mathbf{p}}}
\def\q{{\mathbf{q}}}
\def\k{{\mathbf{k}}}
\def\fp{{p_{\rm 4}}}
\def\fq{{q_{\rm 4}}}
\def\fk{{k_{\rm 4}}}
\def\det{{\mathrm{det}}}
\def\tr{{\mathrm{tr}}}
\def\aka{{\emph{aka}}}
\def\e{\epsilon}
\def\HRULE{{\bigskip\hrule\bigskip}}
\def\be{\begin{equation}}
\def\ee{\end{equation}}
\def\bea{\begin{eqnarray}}
\def\eea{\end{eqnarray}}
There are several reasons to suspect that Lorentz invariance may be
only a low energy symmetry.  This possibility is suggested by the
ultraviolet divergences of local quantum field theory, as well as by
tentative results in various approaches to quantum gravity and string
theory~\cite{Kc,Acea,loopqg1,Alfaro:2000wd,Bruno:2001mw}.  Moreover,
Lorentz symmetry can only ever be verified up to some finite
observationally accessible velocity, which leaves untested an infinite
volume of the supposed symmetry group.

The possibility of Lorentz violation can be studied, without a
particular fundamental theory in hand, by considering its
manifestation in dispersion relations for particles. If rotational
invariance is preserved, it is natural to assume that deviations from
the Lorentz invariant dispersion relation $E^2(p)=m^2 + p^2$ can be
characterized at low energies by an expansion with integral powers of
momentum, $E^2 = m^2 + p^2 + \sum_{n=1}^{\infty} a_n p^n$, where the
$a_n$ are coefficients with mass dimension $2-n$ which might be
positive or negative. [Throughout this letter $p$ denotes the absolute
value of the 3-momentum vector $\p$, and we use units with the low
energy speed of light in vacuum equal to unity.]  Different approaches
to quantum gravity suggest different leading order Lorentz violating
terms. The terms with $n\le4$ have mostly been considered so
far. Observations limit the coefficients $a_{1,2}$ to be extremely
small (see e.g.~\cite{Krev,CG,Stecker:2001vb} and references therein).
In this letter we shall assume they are precisely zero.

The cubic and higher order coefficients have negative mass dimension,
so if the Lorentz violation descends from the Planck scale $M_P=(\hbar
c^5/G)^{1/2}\simeq 1.22 \cdot10^{19}$ GeV, $a_n$ would be expected to
be of order $M_P^{2-n}$. In this case the Lorentz violation is
naturally suppressed, and the lowest order term would dominate all the
higher ones as long as the momentum is less than $M_P$, hence we shall
restrict to a single Lorentz violating term.  We thus consider the
constraints that high energy observations impose on dispersion
relations of the form
\begin{equation}
E_{a}^2 =  p_{a}^2+m^{2}_{a}  +
 \eta_{a} {p_{a}^n}/{M^{n-2}},
 \label{eq:pdr}
\end{equation}
where $a$ labels different fields and $n\ge 3$. We have introduced the
energy scale $M=10^{19}$ GeV $\sim M_P$ so that the coefficients
$\eta_a$ are dimensionless.  If the Lorentz violation comes from
quantum gravity effects, one would expect $\eta_a$ to be within a few
orders of magnitude of unity.  In the absence of a fundamental theory
one has no reason to expect any particular relation between the
coefficients $\eta_{a}$ for different particles, except perhaps that
they should all be of the same order of magnitude.  Since the
dispersion relation (\ref{eq:pdr}) is not Lorentz invariant, it can
only hold in one reference frame.
We assume along with all previous authors that this frame coincides
with that of the cosmic microwave background.  The velocity of the
earth relative to this frame is negligible for our purposes.

Observational consequences of the $\eta$ term in (\ref{eq:pdr}) may
seem out of reach because of the Planck scale suppression.  However
this is not so.  Dispersion relations like (\ref{eq:pdr}) produce
kinematic relations from energy-momentum conservation that differ from
the usual Lorentz invariant case.  As a result reactions can take
place that are normally forbidden, and thresholds for reactions are
modified.  One can expect deviations from standard threshold
kinematics when the the last two terms of (\ref{eq:pdr}) are of
comparable magnitude.  Assuming $\eta$ is of order unity this yields
the condition $p_{\rm dev}\sim (m_{a}^2 M^{n-2})^{1/n}$, which is
$\sim (m_{a}/m_e)^{2/3}\times 10$ TeV for $n=3$ and $\sim
(m_{a}/m_e)^{1/2}\times 10^4$ TeV for $n=4$, where $m_{e}$ is the
electron mass.  Although these energies are currently not achievable
in particle accelerators (except in the case of massive neutrinos
which however are too weakly coupled to provide constraints) they are
in the range of current astrophysical observations.  In fact, it has
been suggested by several
authors~\cite{CG,ACP,Mestres,Bertolami,Kifune,Kluzniak,Aloisio,Protheroe}
(see also \cite{Sigl} and references therein) that we may already be
observing deviations from Lorentz invariance via the possibly missing
Greisen, Zatsepin and Kuzmin (GZK) cut-off on cosmic ray protons with
ultra high energy greater than $7\times 10^{19}$
eV~\cite{GZKobservations}, and the possible overabundance of gamma
rays above 10 TeV from the blazar system Markarian
501~\cite{Protheroe,Aharo2}.  Here we shall mostly not consider the
constraints imposed by asking Lorentz violation to explain these
puzzles.  Instead we restrict our attention to constraints imposed by
consistency with known phenomena (or lack thereof).

\emph{Observational constraints}:
Several studies of observational limits on Lorentz violating
dispersion relations have already been carried
out~\cite{CG,Stecker:2001vb,ACP,Mestres,Bertolami,Kifune,Kluzniak,Aloisio,Kostelecky:2001mb,Gleiser:2001rm,Brustein:2001ik},
with various different assumptions about the coefficients.  Our study
focuses on purely QED interactions involving just photons and
electrons.  We assume $n=3$, since the $n=4$ terms are suppressed by
another inverse power of $M$.  Unlike other studies, no {\it a priori}
relation between the coefficients $\eta_\gamma$ and $\eta_e$ is
assumed, and we combine all the different constraints in order to
determine the allowed region in the parameter plane.  To eliminate the
subscript $a$ we introduce $\xi\, :=\eta_\gamma$, $\eta\, :=\eta_e$,
and $m := m_{e}$.

The modified dispersion relations for photons and electrons in general
allow two processes that are normally kinematically forbidden: vacuum
\v{C}erenkov radiation, $e^-\rightarrow e^-\gamma$, and photon decay,
$\gamma\rightarrow e^+e^-$.  In addition the threshold for photon
annihilation, $\gamma\gamma\rightarrow e^+e^-$, is shifted.  The
vacuum \v{C}erenkov process is extremely efficient, leading to an
energy loss rate that goes like $E^2$ well above threshold.  Similarly
the photon decay rate goes like $E$.  Thus any electron or photon
known to propagate must lie below the corresponding threshold.

We consider constraints that follow from three considerations: (i)
Electrons of energy $\sim 100$ TeV are belived to produce observed
X-ray synchrotron radiation coming from supernova
remnants~\cite{Koyama}, and to also produce multi-TeV photons by
inverse-Compton scattering with these X-rays~\cite{Tanimori98,Naito}.
Assuming these electrons are actually present, vacuum \v{C}erenkov
radiation must not occur up to that energy~\footnote{
The competing energy loss by synchrotron radiation is irrelevant for
this constraint. The rate of energy loss from a particle of energy $E$
due to the vacuum \v{C}erenkov effect goes like $-e^2 E^2$, while that
from synchrotron emission goes like $- e^4 B^2 E^2/m^4$ (using units
where $c=\hbar=1$). For a magnetic field of about one micro Gauss (as
those involved in supernova remnants) the synchrotron emission rate is
$40$ orders of magnitude smaller than the vacuum \v{C}erenkov rate.
}.
(ii) Gamma rays up to $\sim 50$ TeV arrive on earth from the Crab
nebula~\cite{Tanimori}, so photon decay does not occur up to this
energy.  (iii) Cosmic gamma rays are believed to be absorbed in a
manner consistent with photon annihilation off the infrared (IR)
background with the standard threshold~\cite{JStecker01}.  Observation
(iii) is not model independent, so the corresponding constraint is
tentative and subject to future verification.

\emph{Modified kinematics}:
The processes $e^-\rightarrow e^-\gamma$ and $\gamma\rightarrow
e^+e^-$ correspond to the basic QED vertex, but are normally forbidden
by energy-momentum conservation together with the standard dispersion
relations.  When the latter are modified, these processes can be
allowed.  To see this, let us denote the photon 4-momentum by
$k_{4}=(\omega_{k},\k)$, and the electron and positron 4-momenta by
$p_{4}=(E_{p},\p)$ and $q_{4}=(E_{q},\q)$.  For the two reactions
energy-momentum conservation then implies $p_{4}=k_{4}+ q_{4}$ and
$k_{4}=p_{4}+ q_{4}$ respectively.  In both cases, we have
$(p_{4}-k_{4})^2=q_{4}^2$, where the superscript ``2" indicates the
Minkowski squared norm.  Using the Lorentz breaking dispersion
relation Eq.~(\ref{eq:pdr}) this becomes
\begin{equation}
\xi k^3+\eta p^3-\eta q^3 =
2M\left(E_{p}\omega_{k}-pk\cos\theta\right),
\label{Econs}
\end{equation}
where $\theta$ is the angle between $\p$ and $\k$.  In the standard
case the coefficients $\xi$ and $\eta$ are zero and the r.h.s.\ of
Eq.~(\ref{Econs}) is always positive, hence there is no solution.  It
is clear that non-zero $\xi$ and $\eta$ can change this conclusion and
allow these processes.

To derive the observational constraints one needs to determine the
threshold for each process, \ie\ the lowest energy for which the
process occurs.  Assuming monotonicity of all the dispersion relations
(for the relevant momenta $\ll M$) one can show~\cite{lettthr} that
all thresholds for processes with two particle final states occur when
the final momenta are parallel.  Moreover for two particle initial
states the incoming momenta are antiparallel.  This implies that at a
threshold $\theta=0$ in Eq.~(\ref{Econs}) and that in the
corresponding formula for the photon annihilation we shall consider
antiparallel photons and parallel leptons.  These geometries have been
assumed in previous works but to our knowledge they were not shown to
be necessary. In fact they are not necessary if the dispersion
relations are not monotonic. Details concerning the determination of
the thresholds are reported in~\cite{JLM01}.

\emph{Vacuum \v{C}erenkov radiation:}
We find that an electron can emit \v{C}erenkov radiation in the vacuum
if $\eta>0$ or if $\eta<0$ and $\xi<\eta$.  Depending on the values of
the parameters, the threshold configuration can occur with a
zero-energy photon or with a finite energy photon.  These two cases
correspond to the following two threshold relations, respectively:
\begin{eqnarray}
&p_{\rm th}& =
\displaystyle{\left(\frac{m^2M}{2\,\eta}\right)^{1/3}}
\qquad\mbox{for $\eta > 0$ and $\xi\geq-3\eta$},
\label{Cerenkov1} \\
&p_{\rm th}&=
\displaystyle{\left[-\frac{4\,m^2M\left(\xi+\eta\right)}
{\left(\xi-\eta\right)^2}\right]^{1/3}}
\:\:
\begin{array}{ll}
\mbox{for} & \xi<-3\eta<0, \\
&\\
\mbox{or} & \xi<\eta\leq 0.
\end{array}
\label{Cerenkov2}
\end{eqnarray}
The reaction is not allowed in the region where $\xi>\eta$ and
$\eta<0$.  Note that if $\xi=\eta$ only the solution~(\ref{Cerenkov1})
yields a finite threshold.

Electrons of energy $\sim 100$ TeV are indirectly observed via
X-ray synchrotron radiation coming from supernova
remnants~\cite{Koyama}.  Thus for example in the region of the
parameter plane where (\ref{Cerenkov1}) holds we obtain the
constraint $\eta<m^2M/2p_{\rm th}^3\sim 10^{-3}$.

\emph{Photon decay:}
A photon can spontaneously decay into an electron-positron pair
provided $\xi$ is sufficiently great for any given $\eta$.  Contrary
to Lorentz-invariant kinematics of pair creation thresholds, we find
that the two particles of the pair do not always have equal momenta.
Photon decay is allowed above a broken line in the $\eta$--$\xi$
plane given by $\xi=\eta/2$ in the quadrant $\xi,\eta>0$ and by
$\xi=\eta$ in the quadrant $\xi,\eta<0$.  Above this line, the
threshold is given by
\begin{eqnarray}
k_{\rm th}&=& \displaystyle{\left( \frac{8 m^2M}{2\xi-\eta}
\right)^{1/3}
\quad \quad \:\,\,\mbox{for $\xi\geq 0$}},\label{th1}
\\
k_{\rm th}&=&
\displaystyle{\left[ \frac{-8 m^2M\eta}{(\xi-\eta)^2} \right]^{1/3}}
\qquad \;\mbox{for $\eta<\xi<0$.}\label{th2}
\end{eqnarray}
The first relation (\ref{th1}) arises when the electron and
positron momenta are equal at threshold.  The second relation
(\ref{th2}) applies in the case of asymmetric distribution of
momenta.  Note that if $\xi=\eta$, the asymmetric threshold
disappears, leaving just the symmetric one.

The constraint we impose is that the threshold is above 50 TeV, the
highest energy of observed gamma rays from the Crab
nebula~\cite{Tanimori}.  The strength of the constraint is determined
by the smallness of the quantity $m^2M/k^{3}_{\rm max}$.  For $k_{\rm
max}=50$ TeV one gets $m^2M/k^{3}_{\rm max}\approx 0.02$.

\emph{Photon annihilation:}
The standard threshold for a gamma ray to annihilate with an IR
background photon of energy $\e$ is $k_s=m^2/\e$.  In the presence of
dispersion the threshold relations take approximately the same form as
for photon decay, equations (\ref{th1},\ref{th2}), with the
replacement $\xi\rightarrow\xi^{'}$, where $\xi^{'}\equiv
\xi+4\e{M}/k_{\rm th}^2$. (Here we have used the fact that $\e$ is
much smaller than any other scale in the problem.) However, now these
relations correspond respectively to cubic and quartic polynomial
equations for $k_{\rm th}$ (since $\xi'$ is itself a function of
$k_{\rm th}^2$), and the condition that determines whether the
threshold is given by the symmetric (\ref{th1}) or asymmetric
(\ref{th2}) relation is more complicated. The detailed analysis can be
found in~\cite{JLM01}. Here we merely state the result. Rather than
fixing $\eta$, $\xi$ and $\e$ and solving the relations for $k_{\rm
th}$, we fix $\e$ and $k_{\rm th}$ and use the threshold relations to
solve for $\xi$ as a function of $\eta$. When $k_{\rm th}<1.5 k_s$ the
symmetric threshold applies for $\xi'>0$ and the asymmetric one
applies for $\eta<\xi'<0$. When $k_{\rm th}>1.5 k_s$ there is no
symmetric threshold, and the asymmetric one applies only for $\xi$
below the symmetric $k_{\rm th}=1.5 k_s$ line, $\xi=\eta/2 -
(16/27)(\e^3M/m^4)$. In the case $\xi=\eta$ the threshold
configuration is never asymmetric~\cite{JLM01}.

For the observational consequences it is important to
recognize that the threshold shifts are much more
significant at higher energies than at lower energies.
To exhibit this dependence, it is simplest to fix a
gamma ray energy $k$ and to solve for the corresponding
soft photon threshold energy $\e_{\rm th}$. Taking the
ratio with the usual threshold $\e_{{\rm th},0}$,
we find a dependence on $k$ at least as strong as $k^{3/2}$.
Introducing $k_{10}:=k/(10\,\mbox{TeV})$, we have
\begin{eqnarray}
 \frac{\e_{\rm th}}{\e_{{\rm th},0}}&=& 1+
  \frac{(\eta-2\xi)}{20}\,k^{3}_{10}
   \quad \qquad\: \:\:\:\,\mbox{for $\xi'\geq 0$},\label{eth1}\\
 \frac{\e_{\rm th}}{\e_{{\rm th},0}}&=&
  \frac{(\eta-\xi)}{10}\,k^{3}_{10}+\sqrt{-\frac{\eta}{5}\, k^{3}_{10}}
   \quad \mbox{for $\eta<\xi'<0$.}\label{eth2}
\end{eqnarray}

High energy TeV gamma rays from the blazars Markarian 421 and
Markarian 501 have been detected out to 17 TeV and 24 TeV
respectively~\cite{Mrk421,Aharo}.  Although the sources are not
well understood, and the intergalactic IR background is also not
fully known, detailed modeling shows that the data are consistent
with some absorption by photon annihilation off the IR background
(see e.g.~\cite{Mrk421,JStecker01,Aharo2} and references
therein).  However, while the inferred source spectrum for
Markarian
501 is consistent with expectations for energies less than around
10 TeV, above this energy there have been
claims~\cite{Protheroe,Aharo2} that far more photons than
expected are detected.  Nevertheless, recent analysis based on a
more detailed reconstruction of the IR background do not seem to
corroborate this point of view~\cite{JStecker01}.

Due to these uncertainties sharp constraints from photon annihilation
are currently precluded.  Instead, we just determine the range of
parameters $\xi,\eta$ for which the threshold $k_{\rm th}$ lies
between 10 TeV and 20 TeV for an IR photon of energy 0.025 eV with
which a 10 TeV photon would normally be at threshold. Based on current
observations it seems unlikely that the threshold could lie far
outside this range. (It has previously being proposed~\cite{ACP} that
raising this threshold by a factor of two could explain the potential
overabundance of photons over 10 TeV.)  Given the strong energy
dependence of the threshold shift in equations~(\ref{eth1}) and
(\ref{eth2}) this threshold raising would not be obviously in
disagreement with current observations below 10 TeV.

\begin{figure}[htb]
\vbox{ \vskip 8 pt
\centerline{\includegraphics[width=3in]{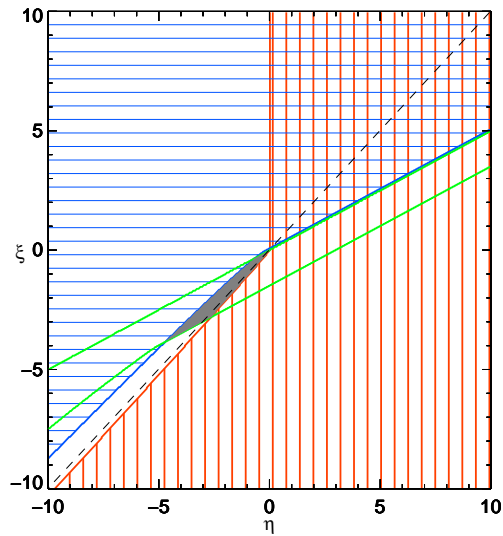}}
\caption{%
Combined constraints on the dimensionless photon and electron
parameters for the case $n=3$ ($\eta,\,\xi=1$ corresponds to the
dimensionful coefficient $M^{-1}=10^{-19}$ GeV${}^{-1}$ of the $n=3$
term in Eq.~(\ref{eq:pdr})).  The regions excluded by the photon decay
and \^{C}erenkov constraints are lined horizontally in blue and
vertically in red respectively.  The region between the two diagonal
green lines corresponds to a threshold between one and two times the
standard threshold (which is $10$ TeV for photon annihilation with an
IR photon of energy 0.025 eV).  The upper green line corresponds to
the unmodified threshold.  The shaded patch is the part of the allowed
region that falls between these photon annihilation thresholds. The
dashed line is $\xi=\eta$.~\label{fig:all}
\smallskip}
}
\end{figure}

\emph{Combined constraints:}
Putting together all the constraints and potential constraints we
obtain the allowed region in the $\eta$--$\xi$ plane (see
Figure~\ref{fig:all}).  The photon decay and \v{C}erenkov constraints
exclude the horizontally and vertically shaded regions, respectively.
The allowed region lies in the lower left quadrant, except for an
exceedingly small sliver near the origin with $0<\eta\lsim 10^{-3}$
and a small triangular region ($-0.16\lsim\eta<0$, $0<\xi\lsim 0.08$)
in the upper left quadrant.  The range of the photon annihilation
threshold previously discussed falls between the two roughly parallel
diagonal lines. The upper diagonal line corresponds to the standard
threshold $k_{\rm s}=10$ TeV and the lower diagonal line to not more
than twice that threshold.  If future observations of the blazar
fluxes and the IR background confirm agreement with standard Lorentz
invariant kinematics, the region allowed by the photon annihilation
constraint will be squeezed toward the upper line ($k_{\rm th}= k_{\rm
s}$).  This would close off {\it all} the available parameter space
except for a region much smaller than unity around the
Lorentz-invariant values $\xi=0=\eta$.

\emph{Conclusions:}
We have shown that astrophysical observations put strong constraints
on the possibility of Lorentz-violating Planck scale cubic
modifications to the electron and photon dispersion relations.  The
constraints arise due to the effect these modifications have on
thresholds for various reactions.  We have also seen that the
threshold configurations with a final state electron-positron pair
sometimes involve unequal momenta for the pair, unlike what occurs for
all Lorentz-invariant decays. This can happen if $\xi\neq\eta$ and
$\xi,\eta<0$.

The allowed region in the $\eta-\xi$ plane includes $\xi=\eta=-1$,
which has been a focus of previous
work~\cite{ACP,Bruno:2001mw,Kifune,Kluzniak}. The negative quadrant
has most of the allowed parameter range. Note that in this quadrant
all group velocities are less than the low energy speed of light.

To further constrain the cubic case will require new observations.
Finding higher energy electrons would not help much, while finding
higher energy undecayed photons would squeeze the allowed region onto
the line $\xi=\eta$.  To shrink the allowed segment of this line using
the reactions we have considered would require observations confirming
the usual threshold for photon annihilation to higher precision.

Perhaps other processes could be used as well.  One might have hoped
that observations comparing the time of flight of photons of different
frequencies from distant sources such as gamma ray bursts and active
galactic nuclei would help constrain the absolute value of $\xi$ (see
e.g.~\cite{Acea,Elltof,schaefer}).  Unfortunately current observations
just yield $|\xi|\lsim 122$ for $n=3$.  This is an interesting
constraint but it is not competitive with the other ones already
considered here. (However the forthcoming Gamma Ray Large Area Space
Telescope (GLAST) mission may provide more stringent constraints of
this type~\cite{Norris99}.)  Another idea is to exploit the fact that
the reaction $\gamma \rightarrow 3 \gamma$ is kinematically allowed
with finite phase space and nonzero amplitude in the presence of
modified dispersion, unlike in the standard case.  This photon decay
channel occurs at {\it all energies} if $\xi>0$, i.e. it has {\it no
threshold}, so it might be thought to provide a very powerful
constraint on positive $\xi$.  Unfortunately, however, the amplitude
for this reaction is far too small to provide any useful
constraint~\cite{JLM01}.

It is interesting to consider the case of the possibly missing GZK
cutoff~\cite{GZKobservations}.  If the cutoff is really missing, it
has been proposed to explain this using Lorentz violating
dispersion~\cite{ACP,CG}.  The relevant protons are at such a high
energy --- over $10^{19}$ eV --- that it takes only tiny Lorentz
violating parameters $\eta_a$ in (\ref{eq:pdr}) to increase the
threshold by an amount of order unity or more. In particular, if one
assumes all coefficients $\eta_a$ are equal, this only requires $\eta$
negative and $|\eta|\gsim m_p^2M^{n-2}/p^n\sim 10^{-38 + 9n}$.  For
$n=3$ this is $10^{-11}$, and for $n=4$ it is still only $10^{-2}$.
Thus for both the $n=3$ and $n=4$ cases only very small values of
$\eta$ are needed to dramatically modify the GZK cutoff, so a shifted
cutoff could be explained by Lorentz-violating constants with our
constraints.  However recent data~\cite{Bahcall:2002wi,FLYeye02}
strongly support the existence of the GZK cutoff at its expected
(Lorentz invariant) value. If this is confirmed, the above analysis
shows that the GZK reaction provides very good constraints for
modifications up to $n=4$~\cite{JLM01}.
%
%

\noindent
\emph{Acknowledgments:} The authors wish to thank G.~Amelino-Camelia,
A.~Celotti and F.~Stecker for helpful discussions.
%
This research was supported in part by NSF grant PHY-9800967.

\end{document}